\documentclass{article}
\usepackage[centertags]{amsmath}

\DeclareMathOperator{\SN}{sn} \DeclareMathOperator{\CN}{cn}
\DeclareMathOperator{\DN}{dn}
\begin{document}

\title{Some notes on elliptic equation method}

\author{ Cheng-shi Liu \\Department of Mathematics\\Daqing petroleum Institute\\Daqing 163318, China
\\Email: chengshiliu-68@126.com\\Tel:86-459-6503476}

 \maketitle

\begin{abstract}
Elliptic equation $(y')^2=a_0+a_2y^2+a_4y^4$ is the foundation of
the elliptic function expansion method of finding exact solutions to
nonlinear differential equation. In some references, some new form
solutions to the elliptic equation have been claimed. In the paper,
we discuss its solutions in detail. By detailed computation, we
prove that those new form solutions can be derived from a very few
known solutions. This means that those  new form solutions are just
new representations of old solutions. From our discussion, some new
identities of the elliptic function can be obtained. In the course
of discussion, we give an example of this kind of formula.


\end{abstract}

\section{Introduction}
 Elliptic equation reads
\begin{equation}
(y')^2=a_0+a_2y^2+a_4y^4
\end{equation}
which is utilized to solve various nonlinear differential
equations(see Refs.[1-8] and the references therein). For example,
Fu et al [1,2] use the solutions of Eq.(1) to give a number of new
kinds of solutions to Sinh-Gordon equation and MKdV equation. It is
well known that its integral form is as follows
\begin{equation}
\int \frac{\mathrm{d}y}{\sqrt{a_0+a_2y^2+a_4y^4}}=\pm(\xi-\xi_0).
\end{equation}
Indeed,  all elliptic function solutions of integral (2) can be
classified by using direct integral method and complete
discrimination system for the third degree polynomial(see, for
example, Refs.[9, 10]). All solutions for the integral of a general
elliptic equation $(y')^2=a_0+a_1y+a_2y^2+a_3y^3+a_4y^4$ can be
obtained (for example, see Ref.[11]). In the paper, for brevity, we
only consider the case of $a_2^2-4a_0a_4>0$. Similarly, other cases
can be easily dealt with. The following result is well known. For
the purpose of completeness and illustration, we list it as follows.
If  $a_4>0$, we rearrange $0, -\frac{1}{b_1}$ and $-\frac{1}{b_2}$,
and denote them as
 $\alpha<\beta<\gamma$. where $b_{1,
2}=\frac{-a_2\pm\sqrt{a_2^2-4a_0a_4}}{2a_4}$ and $b_1\neq b_2\neq0$.
When $\alpha<y^2<\beta$, we have
\begin{equation}
y=\pm\{\alpha+(\beta-\alpha)\SN^2(\sqrt{a_4(\gamma-\alpha)}(\xi-{\xi}_0)),m)\}^{\frac{1}{2}}.
\end{equation}
When $ y^2>\gamma $, we have
\begin{equation}
y=\pm\{\frac{\gamma-\beta\SN^2(\sqrt{a_4(\gamma-\alpha)}(\xi-{\xi}_0)),m)}
{\CN^2(\sqrt{a_4(\gamma-\alpha)}(\xi-{\xi}_0)),m)}\}^{\frac{1}{2}},
\end{equation}
where $ m^2=\frac{\beta-\alpha}{\gamma-\alpha}$.

If $a_4<0$, we rearrange $0, \frac{1}{b_1}$ and $\frac{1}{b_2}$, and
denote them as
 $\alpha<\beta<\gamma$.
 When $\alpha<-w<\beta$, we have
\begin{equation}
y=\pm\{-\alpha-(\beta-\alpha)\SN^2(-\sqrt{-a_4(\gamma-\alpha)}(\xi-{\xi}_0)),m)\}
^{\frac{1}{2}}.
\end{equation}
When $y^2<-\gamma $, we have
\begin{equation}
y=\pm\{-\frac{\gamma-\beta\SN^2(\sqrt{-a_4(\gamma-\alpha)}(\xi-{\xi}_0)),m)}
{\CN^2(\sqrt{-a_4(\gamma-\alpha)}(\xi-{\xi}_0)),m)}\}^{\frac{1}{2}},
\end{equation}
where $ m^2=\frac{\beta-\alpha}{\gamma-\alpha}$.

In the present paper, I discuss in detail the solutions of elliptic
equation (1) and prove that all of those solutions in Ref.[1] (the
results in Ref.[2] can also be dealt with similarly) can be derived
from the solutions (3-6), that is, Fu's solutions just give some new
representations of solutions to the elliptic equation rather than
new solutions. However, we must point out that some new identities
of elliptic functions can be obtained by using these new
representations. In the course of discussion, we give an example of
this kind of formula.

This paper is organized as follows. In section 2, our main results
are given. We verify that all solutions of the elliptic equation in
reference [1] can be represented by the solutions (3-6). At the same
time, a new formula of elliptic function is given. In section 3, we
give a summary.

\section{Transformations of solutions}

For the elliptic equation (1), we verify all of the solutions given
in Ref.[1] are the special cases of the solutions (3-6). We first
list the addition formulae of elliptic functions as follows:
\begin{equation}
\SN(\xi+\eta)=\frac{\SN\xi\CN\eta\DN\eta+\CN\xi\DN\xi\SN\eta}{1-m^2\SN^2\xi\SN^2\eta},
\end{equation}
\begin{equation}
\CN(\xi+\eta)=\frac{\CN\xi\CN\eta-\SN\xi\DN\xi\SN\eta\DN\eta}{1-m^2\SN^2\xi\SN^2\eta},
\end{equation}
\begin{equation}
\DN(\xi+\eta)=\frac{\DN\xi\DN\eta-m^2\SN\xi\CN\xi\SN\eta\CN\eta}{1-m^2\SN^2\xi\SN^2\eta}.
\end{equation}

For brevity, we only discuss the first ten cases in Ref.[1], other
cases can be discussed similarly.

Case 1. $a_0=1-m^2, a_2=2m^2-1, a_4=-m^2$. According to  Eq.(5),
where $\alpha=-1, \beta=0, \gamma=\frac{1-m^2}{m^2}$, we have
\begin{equation}
y=\pm\CN((\xi-\xi_0),m),
\end{equation}
which includes the solution in Eq.(16) of Ref.[1].

Case 2. $a_0=-m^2, a_2=2m^2-1, a_4=1-m^2$. According to  Eq.(4),
where $\alpha=-\frac{m^2}{1-m^2}, \beta=0, \gamma=1$, we have
\begin{equation}
y=\pm\frac{1}{\CN((\xi-\xi_0),m)},
\end{equation}
which includes the solution in Eq.(18) of Ref.[1].

Case 3 and case 4. $a_0=1, a_2=2-m^2, a_4=1-m^2$. According to
Eq.(4), where $\alpha=-\frac{1}{1-m^2}, \beta=-1, \gamma=0$, we have
\begin{equation}
y=\pm\frac{\SN((\xi-\xi_0),m)}{\CN((\xi-\xi_0),m)},
\end{equation}
which includes the solution in Eq.(20) of Ref.[1].

Case 5. $a_0=1, a_2=2m^2-1, a_4=m^2(m^2-1)$. According to  Eq.(5),
where $\alpha=-\frac{1}{1-m^2}, \beta=0, \gamma=\frac{1}{m^2}$, we
have
\begin{equation}
y=\pm\frac{1}{\sqrt{1-m^2}}\CN((\xi-\xi_0),m).
\end{equation}
Since $y(0)=0$ from the solution in Eq.(24) in Ref.[1],  we have
$\CN(\xi_0,m)=0$. Furthermore, we have $\SN^2(\xi_0,m)=1$ and $
\DN^2(\xi_0,m)=1-m^2$. According to the addition formula, we have
\begin{equation}
\CN(\xi-\xi_0)=\frac{-\SN\xi\DN\xi\SN{\xi_0}\DN{\xi_0}}{\DN^2\xi}=\pm\sqrt{1-m^2}
\frac{\SN\xi}{\DN\xi},
\end{equation}
and hence we have
\begin{equation}
y=\pm\frac{\SN\xi}{\DN\xi},
\end{equation}
 which includes the solution in Eq.(24) of Ref.[1].

 Case 6 and case 7. $a_0=1-m^2, a_2=2-m^2, a_4=1$. According to  Eq.(4), where
$\alpha=-1, \beta=m^2-1, \gamma=0$, we have
\begin{equation}
y=\pm\frac{\sqrt{1-m^2}\SN((\xi-\xi_0),m)}{\CN((\xi-\xi_0),m)}.
\end{equation}
Since $y(0)=\infty$ from the solution in Eq.(26) in Ref.[1],  we
have $\CN(\xi_0,m)=0$. Furthermore, we have $\SN^2(\xi_0,m)=1$ and $
\DN^2(\xi_0,m)=1-m^2$. From case 5 we know
\begin{equation}
\CN(\xi-\xi_0)=\frac{\sqrt{1-m^2}\SN\xi}{\DN\xi}.
\end{equation}
In addition, according to the addition formula we have
\begin{equation}
\SN(\xi-\xi_0)=\frac{\CN\xi\DN\xi\SN{\xi_0}}{\DN^2\xi}=\frac{\CN\xi}{\DN\xi}.
\end{equation}
Thus we have
\begin{equation}
y=\pm\frac{\CN\xi}{\SN\xi},
\end{equation}
 which includes the solution in Eq.(26) of Ref.[1].

 Case 8. $a_0=m^2(m^2-1), a_2=2m^2-1, a_4=1$. According to  Eq.(4), where
$\alpha=-m^2, \beta=0, \gamma=1-m^2$, we have
\begin{equation}
y=\pm\frac{\sqrt{1-m^2}}{\CN((\xi-\xi_0),m)}.
\end{equation}
Since $y(0)=\infty$ from the solution in Eq.(31) in Ref.[1],  we
have $\CN(\xi_0,m)=0$. Furthermore we have $\SN^2(\xi_0,m)=1$ and $
\DN^2(\xi_0,m)=1-m^2$. From case 5 we know
\begin{equation}
\CN(\xi-\xi_0)=\frac{\sqrt{1-m^2}\SN\xi}{\DN\xi}.
\end{equation}
Thus we have
\begin{equation}
y=\pm\frac{\DN\xi}{\SN\xi},
\end{equation}
 which includes the solution in Eq.(31) of Ref.[1].

 Case 9 and case 10. $a_0=\frac{1-m^2}{4}, a_2=\frac{1+m^2}{2}, a_4=\frac{1-m^2}{4}.$
 In ref.[1], Fu et al give its two solutions as follows
 \begin{equation}
y=\frac{\CN(\xi,m)}{1\pm\SN(\xi,m)}.
 \end{equation}
 On the other hand, according to Eq.(4), where
$\alpha=-\frac{1+m}{1-m}, \beta=-\frac{1-m}{1+m}, \gamma=0,$ we have
\begin{equation}
y=\pm\frac{\sqrt{\frac{1-m}{1+m}}\SN(\frac{1+m}{2}(\xi-\xi_0),k)}{\CN(\frac{1+m}
{2}(\xi-\xi_0),k)},
\end{equation}
where $ k^2=\frac{4m}{(1+m)^2}$. In order to show the relations of
two kinds of solutions obtained by the above two transformations, we
give the following lemma.

\textbf{Lemma}: If we take $\xi_0$ such that
\begin{equation}
\SN (\frac{(1+m)\xi_0}{2},k)=-\sqrt\frac{1+m}{2},
\end{equation}
then we have
\begin{equation}
\frac{1+\SN(\xi,m)}{\CN(\xi,m)}=\sqrt\frac{1-m}{1+m}\times\frac{\SN(\frac{1+m}
{2}(\xi+\xi_0),k)} {\CN(\frac{1+m}{2}(\xi+\xi_0),k)},
\end{equation}
and
\begin{equation}
\frac{1-\SN(\xi,m)}{\CN(\xi,m)}=\sqrt\frac{1-m}{1+m}\times\frac{\SN(-\frac{1+m}
{2}(\xi+\xi_0),k)} {\CN(-\frac{1+m}{2}(\xi+\xi_0),k)}.
\end{equation}

 \textbf{Proof}: we  consider only the first formula (26). From the solutions (23), we know
 $y(0)=1$, and hence we have
 \begin{equation}
\int_0^\xi \mathrm{d}\xi=\int_1^y
\frac{\mathrm{d}y}{\sqrt{\frac{1-m^2}{4}+\frac{1+m^2}{2}y^2+\frac{1-m^2}{4}y^4}}.
 \end{equation}
 Take the transformation of variable
 \begin{equation}
y=\frac{1-\sin\varphi}{\cos\varphi}.
 \end{equation}
Then
\begin{equation}
-\xi=\int_0^\varphi
\frac{\mathrm{d}\varphi}{\sqrt{1-m^2\sin^2\varphi}},
\end{equation}
and hence we have
\begin{equation}
\frac{1+\SN(\xi,m)}{\CN(\xi,m)}.
\end{equation}
On the other hand, we take another transformation of variable
\begin{equation}
y=\sqrt\frac{1-m}{1+m}\times\frac{\sin\phi}{\cos\phi}.
\end{equation}
Then we have
\begin{equation}
\frac{(1+m)\xi}{2}=\int_{\phi_0}^\phi
\frac{\mathrm{d}\phi}{\sqrt{1-k^2\sin^2\phi}} =\int_0^\phi
\frac{\mathrm{d}\phi}{\sqrt{1-k^2\sin^2\phi}}-\int_0^{\phi_0}
\frac{\mathrm{d}\phi}{\sqrt{1-k^2\sin^2\phi}},
\end{equation}
where $k^2=\frac{4m}{(1+m)^2}, \frac{(1+m)\xi_0}{2}=\int_0^{\phi_0}
\frac{\mathrm{d}\phi}{\sqrt{1-k^2\sin^2\phi}}$ and
$\SN\frac{(1+m)\xi_0}{2}=\sin \phi_0 =-\sqrt\frac{1+m}{2}$. Thus we
have
\begin{equation}
y=\sqrt\frac{1-m}{1+m}\times\frac{\SN(\frac{1+m}{2}(\xi+\xi_0),k)}
{\CN(\frac{1+m}{2}(\xi+\xi_0),k)}.
\end{equation}
The proof is completed.

According to Eq.(25) in the above lemma, if we take $\xi_1=0$ and
$\xi_0$ satisfies $\SN\frac{(1+m)\xi_0}{2}=\sqrt\frac{1+m}{2}$ in
solution (24) and use
$\frac{\CN(\xi,m)}{1-\SN(\xi,m)}=\frac{1+\SN(\xi,m)}{\CN(\xi,m)}$,
then we know solution (23)(take sign "-") and solution (24) (take
sign "+") are the same one. Another case is similar.

Eqs.(26) and (27) can be written as a unit form
\begin{equation}
\frac{1+\SN(\xi-\xi_1,m)}{\CN(\xi-\xi_1,m)}=\sqrt\frac{1-m}{1+m}\times
\frac{\SN(\frac{1+m}{2}(\xi+\xi_0),k)}
{\CN(\frac{1+m}{2}(\xi+\xi_0),k)},
\end{equation}
where $\SN (\frac{(1+m)(\xi_0+\xi_1)}{2},k)=-\sqrt\frac{1+m}{2}$.
This is just a new identity of elliptic functions.

It is easy to see that when $m\rightarrow 0$, we have
$\SN\rightarrow\sin, \CN\rightarrow\cos$. Therefore, if $\sin
\frac{\xi_0}{2}=-\sqrt\frac{1}{2}$,e.g., $\xi_0=-\frac{\pi}{2}$, we
obtain from Eqs.(25-27),
\begin{equation}
\frac{1+\sin\xi}{\cos\xi}=\frac{\sin(\frac{1}{2}(\xi+\xi_0)}
{\cos(\frac{1}{2}(\xi+\xi_0)},
\end{equation}
and
\begin{equation}
\frac{1-\sin\xi}{\cos\xi}=\frac{\sin(-\frac{1}{2}(\xi-\xi_0)}
{\cos(-\frac{1}{2}(\xi-\xi_0)}.
\end{equation}
In fact, these above two equations can be verified easily.

 \textbf{Remark 1}. Indeed, we only use two solutions (4) and (5) to verify
 the first ten cases in Ref.[1].

 \textbf{Remark 2}. Similarly, we can also verify that other twelve
solutions given in Ref.[2] for elliptic equation  can be represented
by the solutions (3-6).

\section{Conclusions}

We verify that all solutions of elliptic equation obtained in
Ref.[1] can be represented by the solutions (3-6). This fact means
that some new form solutions are only new representations of
elliptic functions. Therefore, some solutions claimed in the
references [1,2] are not  novel. However, by using those solutions,
some interesting new formulae of elliptic functions can be obtained.


\begin{thebibliography}{2}
\bibitem{Fu}Fu Z T, Liu S D and Liu S K.  \emph{Commu. Theor. Phys}.
45(2006)55.
\bibitem{F1}Fu Z T, Zhang L, Liu S D and Liu S K.  \emph{Phys. Lett. A}.
325(2004)363.
\bibitem{Fu2}Fu Z T, Liu S D and Liu S K.  \emph{Commun. Theor. Phys}.
40(2003)285.
\bibitem{Fu4}Fu Z T, Chen Z, Liu S K and Liu S D.
 \emph{Commun. Theor.
Phys}. 41(2004)675.
\bibitem{Fu5}Fu Z T, Liu S K and Liu S D. \emph{Commun. Theor. Phys}.
42(2004)343.
\bibitem{ch}Chen H T and Yin H C. \emph{ Communications in nonlinear science and numberical
simulation}. 13(2008)547.
\bibitem{fan}Fan E G.  \emph{Chaos, Solitons
and Fractals}. 16(2003)819.
\bibitem{w}Wang D S, Li H B and Wang J K. \emph{Chaos, Solitons and Fractals}. 2007(in
press,Available online 2 January 2007)
\bibitem{ll}Wang Z X and Guo D R. A survey of special functions. Acadamic Press.
Bejing, 1965.
\bibitem{li2}Liu C S. \emph{Acta. Phys. Sin}. 54(2005)1039(in
Chinese).
\bibitem{Liu4}Liu C S. \emph{Chin. Phys}.
14(2005)1710.

\end{thebibliography}
\end{document}